\documentclass[aps,pre,reprint,floatfix,longbibliography]{revtex4-1}
\pdfoutput=1
\usepackage{graphicx}
\usepackage{amssymb,amsfonts,amsmath}
\usepackage{color}

\newcommand{\papertitle}{Geared topological metamaterials with tunable
mechanical stability}
\newcommand{\Rl}{\mathbf{R}_\text{L}}
\newcommand{\Rt}{\mathbf{R}_\text{T}}
\newcommand{\Nb}{N_\text{b}}
\newcommand{\nb}{n_\text{b}}
\newcommand{\nm}{n_\text{m}}
\newcommand{\nss}{n_\text{ss}}

\begin{document}
\title{\papertitle}
\author{Anne S. Meeussen}
\thanks{These authors contributed equally to this work.}
\affiliation{Instituut-Lorentz, Universiteit Leiden, 2300 RA Leiden, The Netherlands}

\author{Jayson Paulose}
\thanks{These authors contributed equally to this work.}
\affiliation{Instituut-Lorentz, Universiteit Leiden, 2300 RA Leiden, The Netherlands}

\author{Vincenzo Vitelli}
\email{vitelli@lorentz.leidenuniv.nl}
\affiliation{Instituut-Lorentz, Universiteit Leiden, 2300 RA Leiden, The Netherlands}

\begin{abstract}
The classification of materials into insulators and conductors has been shaken
up by the discovery of topological insulators that conduct robustly at the edge
but not in the bulk. In mechanics, designating a material as insulating or
conducting amounts to asking if it is rigid or floppy. Although mechanical
structures that display topological floppy modes have been proposed, they are
all vulnerable to global collapse. Here, we design and build mechanical
metamaterials that are stable and yet capable of harboring protected edge and
bulk modes, analogous to those in electronic topological insulators and Weyl
semimetals. To do so, we exploit gear assemblies that, unlike point masses
connected by springs, incorporate both translational and rotational degrees of
freedom. Global structural stability is achieved by eliminating geometrical
frustration of collective gear rotations extending through the assembly. The
topological robustness of the mechanical modes makes them appealing across
scales from engineered macrostructures to networks of toothed microrotors of
potential use in micro-machines.\end{abstract}

\maketitle

A machine channels power into specific, desired motions. This task is
accomplished through \emph{mechanisms}: rigid parts whose degrees of freedom are
constrained by their assembly so that only the desired motion is allowed. Gears or
cogwheels have been fundamental components of mechanisms since ancient times,
underpinning the earliest known machines~\cite{de1959origin} and driving the
engines of the Industrial Revolution~\cite{musson1969science}.
Moreover, the microfabrication and propulsion of toothed rotors is a key
challenge along the path to feasible 
nanomachines~\cite{DiLeonardo2010,Mirkovic2010,Maggi2015,Williams2015,Erbas-Cakmak2015}.

Mechanisms often underlie the behaviour of mechanical
metamaterials---artificial structures whose mechanical properties arise from the
geometry and arrangement of their building blocks~\cite{Kadic2013,
  Christensen2015}. An appropriately designed
mechanism or soft motion of the repeating unit cell can
translate into an unusual bulk property such as
a negative Poisson ratio~\cite{Lakes1987,Grima2005}, a vanishingly small shear
modulus~\cite{Milton1995,Kadic2012}, a tunable vibrational
response~\cite{Shan2014}, or high-gradient elasticity~\cite{Seppecher2011}.
More generally, the design of a mechanical
metamaterial consists of constraining the available degrees of freedom so
that a desired performance is achieved, much like a simple machine. The degrees
of freedom and associated constraints may arise from simple springs or rigid
beams connected by free hinges, or from more complicated elements such as
flexible beams and blocks of various shapes.  
Despite the strong parallels
between machines and metamaterials, however, the potential for gears as building
blocks of metamaterials and complex mechanisms remains unexplored.

Here, we demonstrate metamaterials that use periodic arrangements of gears and
links to independently tune local flexibility and global stability. These
assemblies of rigid elements harbour zero-energy mechanical modes (free motions
as well as stress-bearing states) that owe their existence to topological
invariants characterizing the mechanical excitations of the periodic structure.
First introduced in isostatic spring networks~\cite{Kane2014,Rocklin2015,Po2016},
topological mechanical modes are insensitive to a wide range of structural
perturbations like their counterparts in electronics~\cite{Hasan2010,Qi2011},
photonics~\cite{Lu2014}, and
acoustics~\cite{Prodan2009,Susstrunk2015,Wang2015a,Wang2015b,Mousavi2015,Khanikaev2015,Nash2015,Huber2016}.
This feature makes them desirable for building metamaterials with robust
mechanical motions and stress-bearing states localized to edges, defects, and
domain walls~\cite{Chen2014,Paulose2015,Paulose2015b,Chen2016,Rocklin2015a}.
While the topology of the excitation spectrum allows us to sculpt spatially
nonuniform mechanical modes, the real-space topology of the link network and the
rotational degrees of freedom are exploited to eliminate uniform deformations
known as Guest-Hutchinson modes~\cite{Guest2003, Kapko2009} that
generically lead to a lack of elastic stability in topological mechanical structures based on
spring networks~\cite{Lubensky2015}.

\begin{figure}[t]
  \centering
  \includegraphics{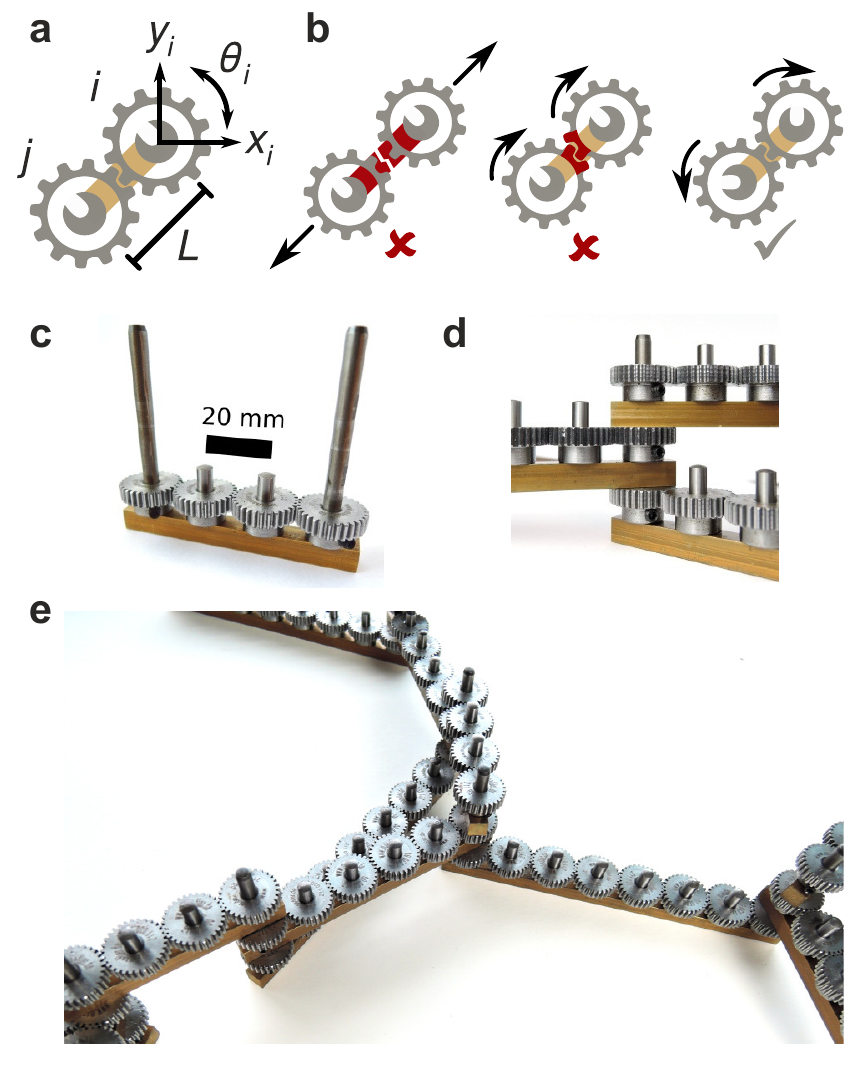}
  \caption{{\bfseries Degrees of freedom and constraints in geared metamaterials.}
    {\bfseries a,} The materials are built up of nodes with three degrees of
    freedom each: two translational ($x_i$, $y_i$) and one rotational ($\theta_i$).
    Nodes $i$ and $j$ are coupled by a single link, represented by the grey bar
    and the pair of gears which rotate together with the nodes.
    {\bfseries b,} The link constrains the distance between the nodes, and
    prevents rotations of the nodes in the same direction. However,
    counterrotations are allowed.
    {\bfseries c,} Realization of such an element using metal parts. The
    degrees of freedom associated with the long axles are coupled by gears
    mounted on a solid bar. The two gears at the ends rotate together with the
    long axles. The shear constraints are enforced using an even
    number of small gears to avoid self-intersections.
    {\bfseries d,} A joint which shares a single set of degrees of freedom,
    associated with the long axle, among three links. The links are stacked
    vertically to avoid self-intersections.
    {\bfseries e,} A unit cell consisting of a specific arrangement of six
    links that, when repeated, would form a geared metamaterial.
  }
  \label{fig:intro}
\end{figure}

\paragraph*{Geared Maxwell networks.}
The fundamental building block of our designs is a pair of gears mounted on a
solid link, as shown in Fig.~\ref{fig:intro}a. Each gear has three independent
degrees of freedom: displacements along the two planar directions, and a
rotation. Each link constrains the distance
between the connected gears, and also prevents motions that involve the gears
sliding against each other (Fig.~\ref{fig:intro}b).
The metamaterials we
envision have a carefully chosen periodic structure, realized in assemblies of metal
links with axles coupled by gears mounted on them (Fig.~\ref{fig:intro}c). Links
of varied lengths are obtained from gears of a single size by chaining even
numbers of gears together. The relevant degrees of freedom, associated with the
long axles at the ends, are shared among bars at junction nodes.
(Fig.~\ref{fig:intro}d).

To describe the zero-energy modes of such a gear network, we first relax the
condition of perfect rigidity and consider links with finite extensibility in
the longitudinal and shear directions. Given displacements $u_{p,i}$ along the
degree of freedom $p \in \{x,y,\theta\}$ of node $i$ (out of $N$ nodes), we
define the longitudinal extension $e_{\text{l},m}$ as the change in length of
link $m$ (out of $\Nb$ links) and the shear extension $e_{\text{s},m}$ as the
transverse deformation of the link ends that is uncompensated by gear
counterrotation. The compatibility matrix $\mathbf{C}$, described in detail in
Appendix~\ref{app:comp}, relates the $2\Nb$-dimensional vector of extensions
$\mathbf{e}=(e_{\text{l},1},e_{\text{s},1},e_{\text{l},2},...)$ to the
$3N$-dimensional displacement vector $\mathbf{u} =
(u_{x,1},u_{y,1},u_{\theta,1},u_{x,2},...)$ via
$\mathbf{C}\mathbf{u}=\mathbf{e}$. Its transpose, the equilibrium matrix,
relates forces and torques acting on the nodes
$\mathbf{f}=(f_{x,1},f_{y,1},\tau_1,f_{x,2},...)$ to a generalized stress vector
$\mathbf{t}=(t_1,s_1,t_2,...)$ of link tensions $t_m$ and shear forces $s_m$ via
$\mathbf{C^T}\mathbf{t}=\mathbf{f}$. \emph{Zero modes} of the structure are
displacements that lie in the null space of $\mathbf{C}$; in a structure where
link extension and shearing are rigidly constrained, these are the only
allowed motions. Members of the null space of $\mathbf{C^T}$ are called \emph{states of
  self-stress}, combinations of tensions and shears that leave all nodes in
equilibrium. The number of zero modes, $\nm$, is related to the number of states
of self-stress $\nss$, via the Maxwell-Calladine index
theorem~\cite{Calladine1978,Kane2014}:
\begin{equation}
  \label{eq:calladine}
  \nm-\nss = 3N-2\Nb.
\end{equation}

Our approach relies on the unique properties of isostatic structures, for which degrees of
freedom and constraints are perfectly balanced---the so-called Maxwell criterion
for structural stability~\cite{Maxwell}. In gear networks, this happens when
$3N=2\Nb$, or equivalently, when the network has average coordination $z =
2\Nb/N = 3$. Eq.~\ref{eq:calladine} highlights the special status of such
\emph{Maxwell lattices}~\cite{Lubensky2015}: zero modes must be accompanied by
an equal number of states of self-stress, and a finite system cut out of an
infinite Maxwell lattice must have zero modes as the perfect balance
is disrupted by the missing links at edges. We note here that the same average
coordination $z=3$ determines the critical point in the jamming of
frictional discs with infinite friction
coefficient~\cite{Shundyak2007,Papanikolaou2013,Henkes2016}. This is not
coincidental: the contact network at jamming also satisfies the Maxwell
criterion~\cite{Liu2010}, and the elements in Fig.~\ref{fig:intro}b replicate
the degrees of freedom and constraints between frictional discs in contact at
the infinite friction limit. Just as the addition of friction changes the
critical coordination at jamming from $z=4$ to $z=3$, the
introduction of rotational degrees of freedom and shear constraints requires
gear networks to be three-coordinated for topological mechanical modes to appear,
unlike topological \emph{spring} networks which must have $z=4$ in two dimensions~\cite{Kane2014,Rocklin2015,Sussman2016a}.  

\begin{figure*}[t]
  \centering
  \includegraphics{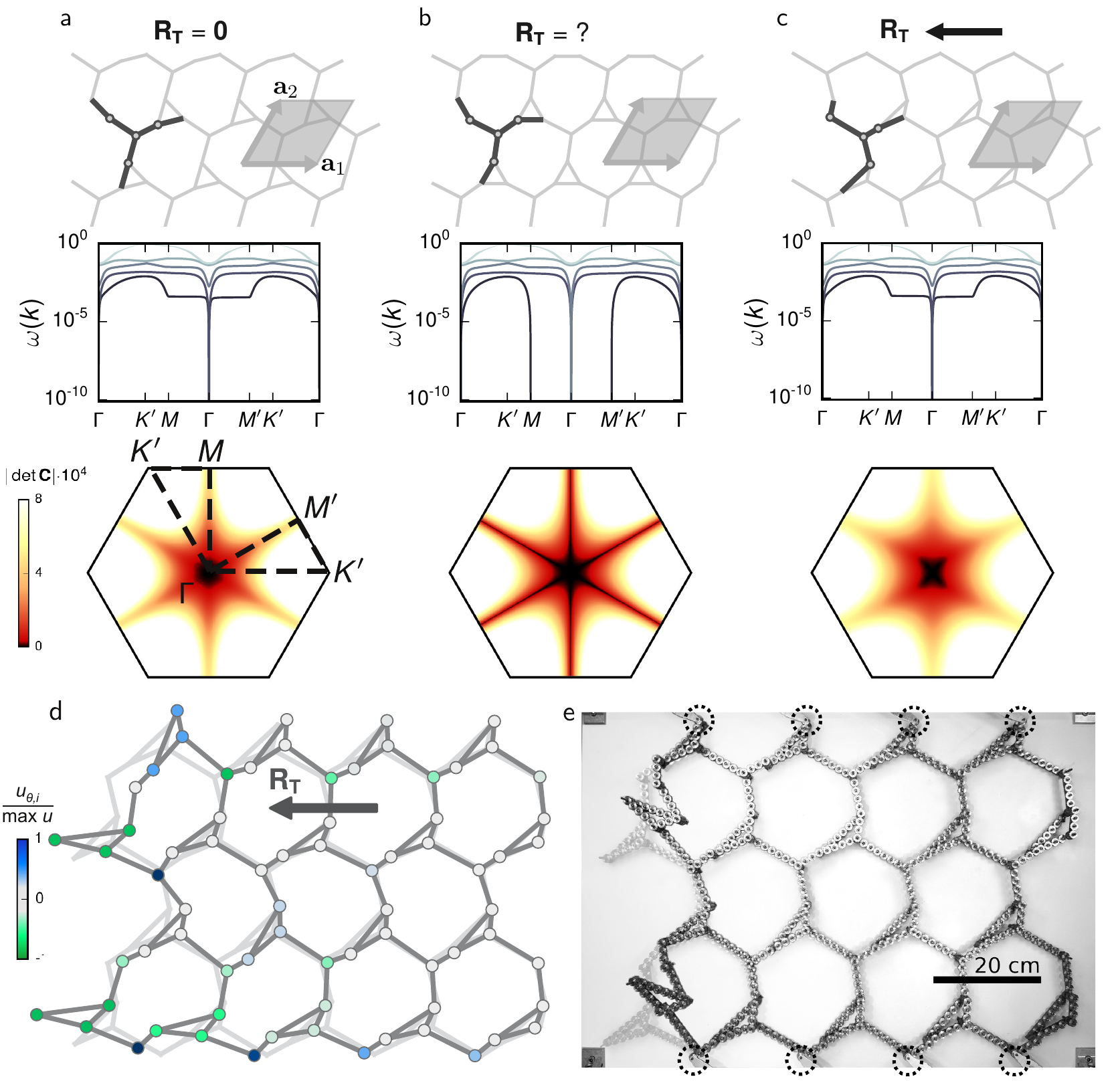}
  \caption{{\bfseries Topologically polarized martini lattices.}
    {\bfseries a--c}, Three members of a single-parameter family of martini lattices.
    The unit cell comprises four nodes (gray dots)  and six unique links (thick
    black lines) and is repeated on a lattice built from primitive vectors
    $\{\mathbf{a}_1,\mathbf{a}_2\}$. Below each lattice are shown the determinant of the
    Fourier-transformed tension-shear compatibility matrix in the Brillouin zone
    (intensity plot) and the lowest six eigenvalues along specific contours (line
    plots).
    The lattices {\bfseries a} and
    {\bfseries c} have a gapped spectrum everywhere except at the
    $\mathrm{\Gamma}$ point ($\mathbf{k}=0)$.
    The topological polarization, calculated as described in the text, is $0$ in
    {\bfseries a} and $-\mathbf{a}_1$ in {\bfseries c}. Lattice {\bfseries b}
    has a gap closing along the line $k_x=0$, and the polarization is not well-defined.
    {\bfseries d,} A strip of the polarized martini lattice with periodic
    boundary conditions along the $y$ direction and free boundaries on the left
    and right. The vectors $\Rl^1$ and $\Rl^2$ quantify the local count of
    zero modes per unit cell along the edge due to missing links. Only the left
    edge localizes zero modes, one of which is visualized by node colours and
    displacements.
    {\bfseries e,} A physical prototype of the framework in {\bfseries a,} made
    out of elements illustrated in Fig.~\ref{fig:intro}. The nodes along the top
    and bottom edges (encircled) are pinned down but free to rotate.
    The left edge has been deformed relative to the unperturbed lattice
    (overlay); no such deformations are possible at the right edge or in the
    interior.
  }
  \label{fig:martini}
\end{figure*}

\paragraph*{Topological characterization.}
For lattices made up of periodically repeating unit cells positioned at linear
combinations of the primitive vectors $\mathbf{a}_i$, the Fourier-transformed extensions and
displacements are related at each wavevector $\mathbf{k}$ in the Brillouin zone (BZ)
via the relation $\mathbf{C(k)u(k)} = \mathbf{e(k)}$. Now, $\mathbf{C(k)}$ is a
$2\nb \times 3n$ complex matrix for a unit cell with $n$ sites and $\nb$ links;
it is a square matrix for Maxwell gear lattices with $2\nb = 3n$. The index theorem,
Eq.~\ref{eq:calladine}, applies at each point in the BZ. Since translations
of the lattice along the $x$ and $y$ directions do not stretch or shear any
links, there are at least two zero modes at $\mathbf{k}=0$, with two
accompanying states of self-stress. However, for certain
unit cell geometries, $\mathbf{C(k)}$ has no zero eigenvalues away from
$\mathbf{k}=0$; we call this a \emph{gapped} spectrum. Kane and Lubensky have
recently shown~\cite{Kane2014} that square compatibility matrices with
gapped spectra can be classified by a set of $d$ topological indices for
$d$-dimensional lattices: the integer winding numbers $\{n_1,...,n_d\}$ of
the phase of $\det \mathbf{C(k)} = |\det \mathbf{C(k)}|e^{i\phi(\mathbf{k})}$,
\begin{equation} \label{eqn_windingnumbers}
    n_i=- \frac{1}{2\pi}\oint_{C_i}d\mathbf{k}\cdot\nabla_\mathbf{k}\,\phi(\mathbf{k})
\end{equation}
along the $d$ topologically distinct cycles $C_i$ of the BZ
parallel to the reciprocal lattice directions $\mathbf{b}_i$ 
defined by $\mathbf{a}_i\cdot\mathbf{b}_j=2\pi\delta_{ij}$. The topological
nature of the $n_i$ is reflected in the fact that changing their values requires
the spectral gap to close.
The indices can be
used to define a lattice vector $\Rt = \sum_i n_i\mathbf{a}_i$, called
the \emph{topological polarization} because it can be interpreted as a transfer
of degrees of freedom along its direction as we will demonstrate. 

To find Maxwell gear networks  with a spectral gap and
nontrivial topology (nonzero winding numbers of $\mathbf{C(k)}$), we consider modifications of the
martini lattice~\cite{Scullard2006}, which has four sites and six links per unit
cell on a regular hexagonal Bravais lattice  (see Appendix~\ref{app:unitcells} for
descriptions of all lattices used in this work).
By changing node positions and link lengths in the unit cell without affecting
link connectivity , we find a range of distorted martini lattices with gapped
spectra. Fig.~\ref{fig:martini}a and c show two examples. The
lattice in Fig.~\ref{fig:martini}a has $n_1=n_2=0$ whereas the lattice in
Fig.~\ref{fig:martini}b has $n_1=-1$ and $n_2=0$. As required, a one-parameter
family of unit cells that connects the two lattices has a member with a gap
closing in the spectrum of $\mathbf{C(k)}$, shown in Fig.~\ref{fig:martini}b.
The compatibility matrix of this lattice has a line of zero eigenvalues
extending through the BZ along $k_x=0$, so the winding numbers of
$\det \mathbf{C(k)}$ become ill-defined.

\paragraph*{Edge mode polarization.}
The winding numbers influence the count of zero modes at lattice
edges~\cite{Kane2014}. Eq.~\ref{eq:calladine} dictates that a system cut
out of an infinite Maxwell lattice will have an excess of zero modes equal to the number
of deficient links, which is proportional to the length of the cut edges. When
the compatibility matrix has a gapped spectrum, these zero modes cannot
penetrate into the bulk (except for the two rigid-body translations at
$\mathbf{k}=0$) and must therefore be localized at free
surfaces~\cite{Sun2012,Lubensky2015}. As detailed in Appendix~\ref{app:pol}, the
count of zero modes localized at a free edge has two contributions: a
\emph{local count} $\nu_\text{L}$, which depends on the details of the edge
termination, and a \emph{topological count} $\nu_\text{T}$, which depends solely
on the orientation of the edge normal relative to the bulk lattice polarization
and is independent of local properties at the edge~\cite{Kane2014}. For example,
the strip shown in Fig.~\ref{fig:martini}d (with periodic
boundary conditions along the vertical direction) has an identical number of
missing constraints on the left and the right free edges, and the local count is
$\nu_\text{L}=2$ per unit cell on both edges. However, 
the topological contributions are equal and opposite: $\nu_\text{T}=+2$ and $-2$
on the left and right edges respectively. The net count
$\nu_\text{L}+\nu_\text{T}$ is four per edge unit cell on the left edge, and
zero on the right edge. The topological polarization $\Rt$ effectively transfers
zero modes along its orientation, from the right edge to the left. A numerical calculation
(Appendix~\ref{app:edgenumerics}) confirms that all
non-translational zero modes are exponentially localized to the left edge, even
for small separations between the left and right edges. The
displacements and rotations associated with one such mode are visualized in
Fig.~\ref{fig:martini}d. 

To demonstrate the consequences of the topological bias in zero-mode
distribution, we built a mechanical prototype of the polarized lattice with
two free edges that run perpendicular to $\Rt$, as shown
in Fig.~\ref{fig:martini}e. The topological polarization leads to a
drastic asymmetry in stiffness between the left and right edges of the
prototype, built from the metallic units shown in Fig.~\ref{fig:intro}. As shown in
Supplementary Movie 1 and Fig.~\ref{fig:martini}e, the right edge and the bulk
of the prototype are rigid under manual probing (aside from mechanical play), but the left edge
can be deformed significantly from its original configuration.
The topological character of
the lattice governs the mechanical response of the prototype beyond the linear
regime addressed by the mode calculation, giving rise to 
flexibility along a specific edge of an otherwise rigid structure. The
polarization can also induce topological modes and states of self-stress at
appropriately oriented domain walls separating lattices of different
orientation, even if no bonds have been cut (Appendix~\ref{app:domainwall}).

\begin{figure*}
  \centering
  \includegraphics{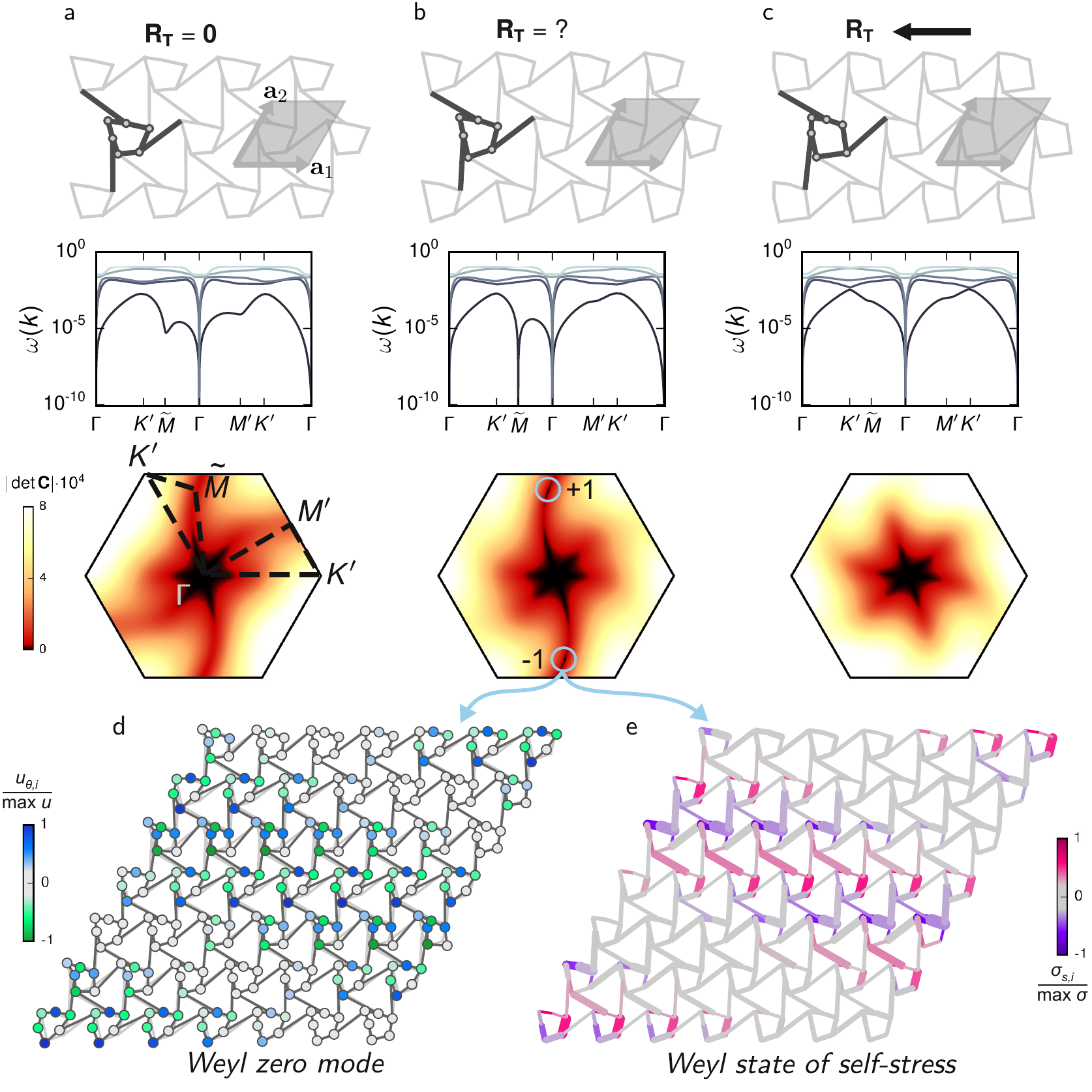}
  \caption{{\bfseries Topological hexachiral lattices.} {\bfseries a--c}, Three
    members of a single-parameter family of hexachiral gear lattices. The
    symmetric unit cell comprises six nodes (gray dots) and nine unique links
    (thick black lines). Below each lattice are shown the determinant of the
    Fourier-transformed tension-shear compatibility matrix in the BZ
    (intensity plot) and the lowest six eigenvalues along specific contours (line plots).
    The lattices {\bfseries a} and {\bfseries c} have a gapped spectrum
    everywhere except at $\mathbf{k}=0$. The topological polarization,
    calculated as described in the text, is $0$ in {\bfseries a} and
    $-\mathbf{a}_1$ in {\bfseries c}. Lattices {\bfseries a} and {\bfseries c}
    are separated by a family of lattices with a pair of Weyl points at which a
    zero eigenstate exists. One such
    lattice is shown in {\bfseries b} with the Weyl points at
    $\mathbf{k}=\pm(-0.295/a,3.001/a)$ encircled. The winding numbers
    $n_\text{w}=\pm 1$ are indicated.
    {\bfseries d--e,} Visualization of the Weyl zero mode ({\bfseries d}) and
    state of self-stress ({\bfseries e}) at the lower Weyl point. The periodic
    variation of the modes is a consequence of the finite wavevector of the
    $\widetilde{M}$-point at which the gap closes.
  }
  \label{fig:weyllattices}
\end{figure*}

\paragraph*{Topological Weyl modes.}
In addition to modes localized at edges, Maxwell gear
networks also support topologically protected 
zero modes and states of self-stress that extend into the bulk of a
sample. These states, termed Weyl modes~\cite{Rocklin2015}, are associated with
gap closings at isolated ``Weyl points'' in the BZ away from $\mathbf{k}=0$
with nonzero windings $n_\text{w}=-\oint_C d\mathbf{k}\cdot\nabla_\mathbf{k}
\phi(\mathbf{k})/(2\pi)$, where $C$ is a contour in momentum space encircling the
point at which $\det \mathbf{C(k)}=|\det \mathbf{C(k)}| e^{i\phi(\mathbf{k})}
=0$. Time-reversal symmetry ensures that
$\mathbf{C(k)}$ and $\mathbf{C(-k)}$ share the same eigenvalues, which means
that Weyl points always occur in pairs with equal and opposite windings. Weyl
points and their associated Weyl modes can only be removed if pairs with opposite
windings annihilate at high-symmetry points in the BZ. 
To realize such modes in geared metamaterials, we consider networks
based on the hexachiral lattice~\cite{Prall1997} which has the same primitive
vectors as the martini lattice and a 6-node, 9-link unit cell which forms
hexagonal and triangular plaquettes (Figure~\ref{fig:martinihex}b). The compatibility matrix
of the regular hexachiral lattice has an ungapped spectrum, but distorting the
plaquettes of the hexachiral lattice into irregular six-sided polygons can open
up a spectral gap everywhere except at isolated points. An example is
shown in Fig.~\ref{fig:weyllattices}b, which has gap closings at $\mathbf{k}=\pm
(-0.3/a,3.0/a)$ with nonzero winding numbers $n_\text{w}=\pm 1$. The periodic
modulation of the corresponding topological Weyl modes is visualized in
Fig.~\ref{fig:weyllattices}d--e.

The structure shown in Fig.~\ref{fig:weyllattices}b is a member of a one-parameter
family of distorted hexachiral lattices which smoothly interpolates between two
lattices with gapped spectra and differing topological polarizations, shown in
Figs.~\ref{fig:weyllattices}a and c. Upon following this family of lattices, as
is done in Supplementary Movie 2, the gap first closes at the point
$\mathbf{k}=(0,2\pi/\sqrt{3}a)$ on the boundary of the BZ, which
then splits into a pair of Weyl points with opposite winding numbers
$n_\text{w}=\pm 1$. The Weyl points sweep through the BZ, moving toward
$\mathbf{k}=0$, where they annihilate.  The change in
the value of $\Rt$ between the gapped spectra in Figs~\ref{fig:weyllattices}a and c is
effected by this traversal of Weyl points with nontrivial winding numbers across
the BZ.

\begin{figure*}
  \centering
  \includegraphics{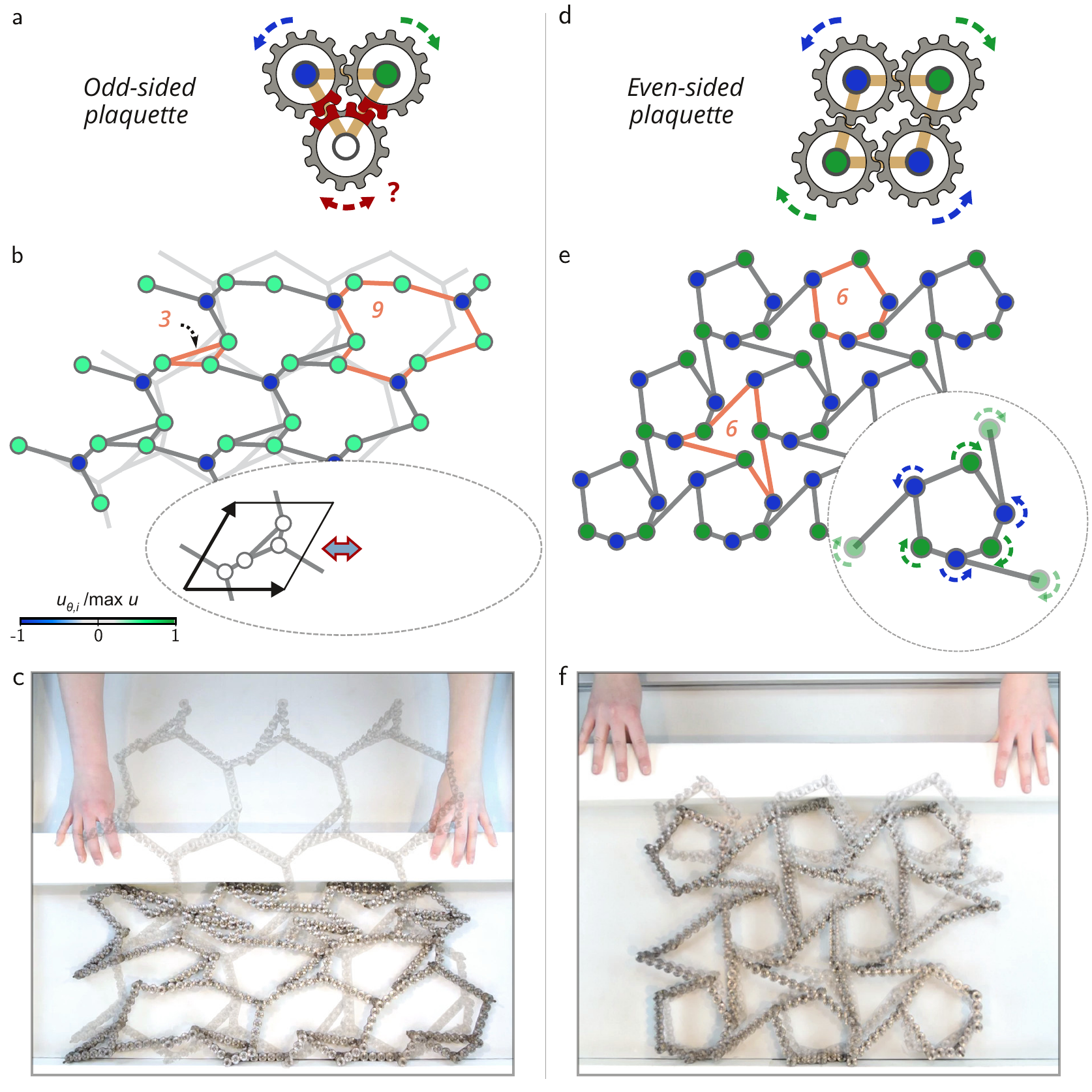}
  \caption{{\bfseries Frustration and global deformations in martini and hexachiral
      lattices.} Maxwell gear lattices with gapped spectra have a global zero-energy
    mode that differs qualitatively in the martini and hexachiral designs.
    {\bfseries a,} When an odd number of links is connected in a loop,
    rotation of gears without node displacements is frustrated.
    {\bfseries b,} The martini lattices have only odd-sided plaquettes with
    either 3 or 9 links (highlighted). Therefore, the zero-energy mode
    (visualized in the overlay) is a Guest-Hutchinson mode that requires point
    displacements as well as a distortion of the unit cell.
    The zoom shows the zero-energy deformation of the unit cell,
    including the change in primitive lattice vectors.
    {\bfseries c,} A prototype martini lattice with free boundaries cannot resist
    vertical compression (horizontal bar being pushed down) because of the
    Guest-Hutchinson mode (the pre-compression configuration is shown in overlay).  
    {\bfseries d,} Gears on an even-sided loop of links can counterrotate freely
    without node displacements.
    {\bfseries e,} A hexachiral lattice consists solely of six-sided plaquettes
    (highlighted).  It can be divided into two sublattices, with all gears
    rotating counterclockwise (blue) on one sublattice and clockwise (green) by
    the same amount on the other without frustration. This is
    the required global zero-energy mode of the periodic system, which does not
    displace any nodes.
    {\bfseries f,} The absence of a unit-cell-shape changing mode imparts
    stability to the hexachiral lattice, which largely maintains its shape under uniform
    compression.  
  }
  \label{fig:guestmodes}
\end{figure*}

\paragraph*{Guest-Hutchinson modes and global stability.}
Periodic Maxwell lattices are perfectly balanced when degrees of freedom and
constraints within the unit cell are considered. However, to build a periodic
lattice, the primitive vectors $\mathbf{a}_i$ determining the relative placement
of unit cells must also be specified. The two components for each
primitive vector add up to four additional degrees of freedom.
According to the index theorem, Eq.~\ref{eq:calladine}, the periodic
lattice therefore has four excess zero modes when primitive vector changes are allowed.
Three of these are the rigid-body transformations(two translations and a
rotation) that do not deform any links. The remaining mode might either be
decoupled completely from lattice displacements, or correspond to an affine
deformation that globally expands or shrinks the lattice termed the
Guest-Hutchinson mode~\cite{Guest2003}. All Maxwell spring lattices have a
Guest-Hutchinson mode which, being a zero-energy affine deformation, leads to
global collapse of the lattice under some external
stresses~\cite{Guest2003,Kapko2009,Lubensky2015}. The Guest-Hutchinson mode
makes topological spring networks elastically unstable,
but can be exploited to globally transform their topological
properties~\cite{Rocklin2015a}.  

The martini and hexachiral gear lattices differ drastically in their
global mechanical stability when affine lattice deformations are taken into
account. In the martini lattices, gear rotations are not allowed without node displacements since they
are frustrated on the odd-sided plaquettes making up the lattice
(Fig.~\ref{fig:guestmodes}a), and the nontrivial zero mode of the augmented
compatibility matrix involves node displacements and a change in primitive vectors
depicted in Fig.~\ref{fig:guestmodes}b.
As a result, the martini gear lattice is unstable against uniform
compressions along the vertical direction (Supplementary Movie 3 and
Fig.~\ref{fig:guestmodes}c).

By contrast, hexachiral gear lattices avoid the nonlinear collapse mode by taking advantage
of rotational degrees of freedom. Unlike the martini lattices, they are
comprised solely of even-sided plaquettes which support gear counterrotations
without node displacements. The two-color theorem of graphs~\cite{Conway2010}
ensures that such lattices are bipartite, i.e. divisible into two sublattices so
that no two nodes in the same sublattice are linked. As a result, they
harbour a global zero-energy mode at $\mathbf{k}=0$ 
in which gears on the two sublattices counterrotate without
hindrance~\cite{Gordon1994}, as shown in  Fig.~\ref{fig:guestmodes}e.
These collective rotations absorb the Guest-Hutchinson mode, eliminating the
global collapse mechanism. Consequently, gapped hexachiral gear lattices
are stable against external loads (Fig.~\ref{fig:guestmodes}f). 
Note that the lattice is 
unstable against \emph{torques} applied on individual gears (which may drive the
global counterrotation mode without resistance), but this will not lead to shape-changing
deformations. 

\paragraph*{Towards geared metamaterials.}
We now comment on practical considerations in implementing the proposed designs.
The topological mechanisms are independent of the size of the building blocks,
as long as the relevant degrees of freedom and constraints
(Fig.~\ref{fig:intro}) are incorporated. At architectural or tabletop scales,
the designs may be realized by assembling beams and gears manually as
demonstrated by our prototypes, but the number of repeating units is limited by
practical constraints on time and effort. The number of gears needed per unit
cell could be dramatically reduced if gears of multiple diameters are used. To
increase the number of unit cells, however, the assembly process itself must be
automated. Modern additive manufacturing techniques allow multicomponent
assemblies with free~\cite{Cali2012} or lubricated~\cite{Maccurdy2016} joints to
be 3D-printed in one shot, providing a platform to scale up the
designs here to structures with hundreds or thousands of unit cells while
bringing down the size of the individual units to microscopic scales.

Construction may be simplified by using gear-free elements which approximate the
required distance and shear constraints. In Appendix~\ref{app:alternative}, we
describe an assembly of four freely-hinged beams which reproduces the
constraints of the gear link to linear order in node displacements and
rotations, and therefore generates the same entries in the compatibility matrix.
A structure built out of these elements would display the same linear
excitations (including zero-energy modes) as the gear network with fewer
moving parts, but its nonlinear response at large deformations would differ. The
hinged beams may be implemented as continuous elements with slender flexible
joints~\cite{Kadic2012}, enabling fabrication via standard additive
manufacturing techniques over a wide range of length scales.

Regardless of fabrication technique, real structural elements are liable to
incorporate additional degrees of freedom (e.g. mechanical backlash or play) and
constraints (e.g. finite friction of gear shafts) beyond those considered by the
model. As is true for any metamaterial design, the predicted behaviour will be
observed as long as the desired degrees of freedom and constraints dominate the
spurious ones for relevant loads. For instance, if the components are
sufficiently rigid, the torque needed to overcome shaft friction is associated
with a negligibly small deformation of the gears compared to the desired motion.
Our prototypes demonstrate that this separation is achievable in practice.
Unconstrained play is clearly distinguishable from true mechanisms which involve
coordinated node motion and gear rotation (Supplementary Movie 1). Manually
actuating the mechanisms requires some effort to overcome substrate and
gear-shaft friction, but the forces involved are negligible compared to those
needed to significantly deform the metal links.

The topological nature of the zero-energy modes ensures their protection
against a large range of structural imperfections. For example, the bias in edge
mechanisms is predicted by the topological polarization
(Fig.~\ref{fig:martini}d--e) for a periodic lattice in the bulk, but is guaranteed to persist
for any perturbation of the interior points as long as the perturbation does not
close the bulk phonon gap. Such a gap closing would correspond to the appearance
of zero-energy modes that extend from the left edge to the right edge of the
system, requiring large coordinated global changes of the unit cells. As long as
the chosen unit cell is far from singular gap-closing configurations such as the
lattice in Fig.~\ref{fig:martini}b, perfect crystalline order is not required for the
topological bias in zero modes to persist. Similarly, local defects in the
structure such as slipped, jammed, or missing gears may introduce localized
zero modes or states of self-stress, but their influence on localized
boundary modes is exponentially small in their distance from the boundary.
Defects on or near the left boundary in Fig.~\ref{fig:martini}d may add or
remove individual zero modes, but cannot change the overall flexibility of the
edge which arises from a macroscopic number (proportional to the length of the
edge) of topological modes. 

\paragraph*{Conclusion.}
We have designed and built topological geared metamaterials in
which translational and rotational degrees of freedom are constrained in a
geometrically controlled manner giving rise to tunable mechanical properties.
Gear lattices satisfying the Maxwell criterion for marginal rigidity support
zero modes and stress states of topological origin like their spring lattice
counterparts, but with the additional ability to independently tune the global
elastic stability. The absence of a zero-energy collapse mode in certain
unfrustrated gear lattices, which is a direct consequence of the rotational
degrees of freedom, paves the way towards building topological mechanical
structures that maintain their shape without requiring extra constraints such as
pinning to a static background. The key insight, that expanded degrees of
freedom and interactions that go beyond harmonic springs open up new
possibilities for topological mechanics, is applicable to a wide range of
systems including colloidal or molecular assemblies that interact via
non-central forces.

\begin{acknowledgments}
  We thank  the Leiden University Fine Mechanics
  Department for technical support, and Bryan Chen and Anton Souslov for useful discussions. This
  work was funded by FOM, a Delta ITP Zwaartekracht grant, and a VIDI grant from NWO.   
\end{acknowledgments}

\appendix
\section{Compatibility matrix for gear elements} \label{app:comp}
The compatibility matrix $\mathbf{C}$ of a gear lattice is built up of contributions from
individual links. Each constraint adds a row to the matrix and each degree of
freedom is assigned to a column. The contribution of link $m$ of
length $L$ which connects  nodes $i,j$ is given by~\cite{Hutchinson2005}:
\begin{equation}\label{eq:cblockrmat}
  \begin{pmatrix}
    e_{\text{l},m} \\
    e_{\text{s},m} \\
  \end{pmatrix}	
  =
  \begin{pmatrix}
    \frac{L_x}{L} & \frac{L_y}{L} & 0 & -\frac{L_x}{L} & -\frac{L_y}{L} & 0\\
    -\frac{L_y}{L}  & \frac{L_x}{L}  & \frac{L}{2} & \frac{L_y}{L}  & -\frac{L_x}{L} & \frac{L}{2} \\
  \end{pmatrix}
  \begin{pmatrix}
    u_{x,i} \\
    u_{y,i} \\
    u_{\theta,i} \\
    u_{x,j} \\
    u_{y,j} \\
    u_{\theta,j} \\		
  \end{pmatrix}		
\end{equation}				
Here, $L_x$ and $L_y$ are the projections of the link's length along $\hat{x}$ and $\hat{y}$; $e_{\text{l},m}$ and $e_{\text{s},m}$ represent the tension and shear strains on the link; and $u_{x,i}, u_{y,i}$ and $u_{\theta,i}$ denote the translational and rotational displacements of the $i^{\mathrm{th}}$ node.

Therefore, each link in an assembly provides two rows to the compatibility
matrix. The number of columns expands to the degrees of freedom in the
assembly, with zero entries for degrees of freedom unassociated with
the link that contributes a particular row.

For periodic lattices, the Fourier-transformed compatibility matrix
$\mathbf{C(k)}$ has one row/column for each constraint/degree of freedom in the
unit cell. Row entries for links that connect nodes in different unit cells
separated by a Bravais lattice vector $R$ acquire a complex phase factor
$e^{i\mathbf{k \cdot R}}$.  

\section{Unit cell definitions for periodic lattices} \label{app:unitcells}
\begin{figure}
  \centering
  \includegraphics{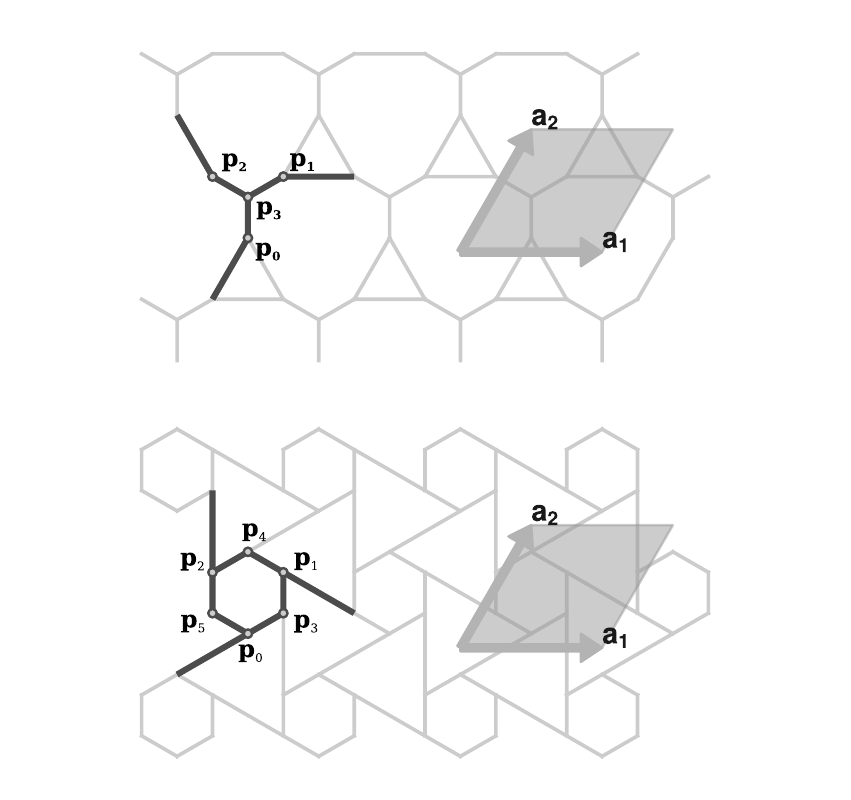}
  \caption{{\bfseries Regular martini (top) and hexachiral (bottom) lattices.}
    Symmetric unit cells are shown with points labelled.}
  \label{fig:martinihex}
\end{figure}

The three-coordinated Maxwell lattices we have considered are based on the
martini lattice, originally developed to study percolation~\cite{Scullard2006},
and the hexachiral lattice, which formed the basis of an early negative Poisson
ratio metamaterial~\cite{Prall1997}. Fig.~\ref{fig:martinihex} shows the unit
cells for the regular martini and hexachiral lattices. The lattices use the same
underlying hexagonal Bravais lattice, with primitive vectors $\mathbf{a}_1=a
\hat{x}$, $\mathbf{a}_2 = (a/2)\hat{x}+(\sqrt{3}a/2)\hat{y}$. The martini
lattice has four nodes and six unique links per unit cell, whereas the
hexachiral unit cell has six nodes and nine unique links.

The lattices studied in the main text are completely defined by specifying the
positions of the nodes labelled in Fig.~\ref{fig:martinihex}. Tables
\ref{tab:martiniunitcells} and \ref{tab:hexunitcells} provide definitions of the
distorted martini and hexachiral lattices respectively.

\begin{table*}
  \centering
  \begin{tabular}{|c|c|}
    \hline
    {\bfseries Lattice} & {\bfseries Node positions} \\
    \hline
    A & $\mathbf{p}_{0}=(-0.05,-0.38)$, $\mathbf{p}_{1}=(0.35,0.15)$, $\mathbf{p}_{2}=(-0.30,0.23)$, $\mathbf{p}_{3}=(0.05,-0.00)$\\
    B & $\mathbf{p}_{0}=(0.00,-0.38)$, $\mathbf{p}_{1}=(0.33,0.19)$,
    $\mathbf{p}_{2}=(-0.33,0.19)$, $\mathbf{p}_{3}=(0.05,-0.00)$\\
    C & $\mathbf{p}_{0}=(0.15,-0.38)$, $\mathbf{p}_{1}=(0.25,0.11)$,
    $\mathbf{p}_{2}=(-0.40,0.27)$, $\mathbf{p}_{3}=(0.05,-0.00)$\\
    \hline
  \end{tabular}
  \caption{{\bfseries Unit cell descriptions of distorted martini lattices.}
    Labels refer to the lattices in Fig.~2 of the main text. The position of
  the nodes (labelled in Fig.~\ref{fig:martinihex}a) is specified in units of the
  lattice constant $a$; the primitive vectors and link connectivity are unchanged.}
  \label{tab:martiniunitcells}
\end{table*}

\begin{table*}
  \centering
  \begin{tabular}{|c|c|}
    \hline
    {\bfseries Lattice} & {\bfseries Node positions} \\
    \hline
    A & $\mathbf{p}_{0}=(-0.14,-0.04)$, $\mathbf{p}_{1}=(0.22, -0.22)$, $\mathbf{p}_{2}=(0.04, 0.17)$, $\mathbf{p}_{3}=(-0.07, -0.25)$, $\mathbf{p}_{4}=(0.36, 0.10)$, $\mathbf{p}_{5}=(-0.20, 0.20)$ \\
    B & $\mathbf{p}_{0}=(-0.15, -0.04)$, $\mathbf{p}_{1}=(0.23, -0.23)$, $\mathbf{p}_{2}=(0.04, 0.17)$, $\mathbf{p}_{3}=(-0.09, -0.33)$, $\mathbf{p}_{4}=(0.31, 0.08)$, $\mathbf{p}_{5}=(-0.20, 0.20)$ \\
    C & $\mathbf{p}_{0}=(-0.17, -0.04)$, $\mathbf{p}_{1}=(0.31, -0.31)$,
    $\mathbf{p}_{2}=(0.04, 0.14)$, $\mathbf{p}_{3}=(-0.10, -0.36)$,
    $\mathbf{p}_{4}=(0.25, 0.07)$, $\mathbf{p}_{5}=(-0.16, 0.16)$ \\
    D &
    $\mathbf{p}_{0}=(-0.44,-0.20)$,$\mathbf{p}_{1}=(0.07,-0.21)$,$\mathbf{p}_{2}=(-0.08,0.42)$,
    $\mathbf{p}_{3}=(-0.19,-0.29)$,$\mathbf{p}_{4}=(0.17,0.05)$,$\mathbf{p}_{5}=(-0.49,0.24)$ \\
    \hline
  \end{tabular}
  \caption{{\bfseries Unit cell descriptions of distorted hexachiral lattices.}
    Labels A--C refer to the lattices in Fig.~3 of the main text; D is the
    lattice used in Fig.~4 of the main text. The position of
  the nodes (labelled in Fig.~\ref{fig:martinihex}b) is specified in units of the
  lattice constant $a$; the primitive vectors and link connectivity are unchanged.}
  \label{tab:hexunitcells}
\end{table*}

\section{Topological polarization and zero-energy edge modes} \label{app:pol}

The formalism for counting the number of net degrees of freedom (i.e. number of
zero modes minus the number of states of self-stress) localized to a particular
edge of a Maxwell lattice with a gapped spectrum was developed in
Ref.~\onlinecite{Kane2014} (see also Ref.~\onlinecite{Lubensky2015}). In this
Appendix, we summarize the main results in the context of gear lattices.

As explained in the main text, free edges of Maxwell lattices are associated
with zero-energy modes. There are two contributions to the number of zero modes
per repeating unit along the edge. The first is a local count of the missing
links along the edge, whose computation can be explained using an electrostatic
analogy. Upon associating ``charges'' of +3 (for the three degrees of freedom)
to each node and -2 (for the two constraints) to the centre of each link, a
Maxwell lattice has a charge-neutral unit cell, but a dipole
moment~\cite{Kane2014} can be defined as
\begin{equation}
  \label{eq:dipolemoment}
  \mathbf{d} = 3\sum_i \mathbf{r}_i - 2\sum_m \mathbf{r}_m,
\end{equation}
where $\mathbf{r}_i$ and $\mathbf{r}_m$ are respectively the position vectors of
node $i$ and link $m$ within the unit cell. At a particular edge, a unit cell
must be chosen which is compatible with the edge termination (with no dangling
bonds), leading to a nonzero edge-specific dipole moment $\Rl$. The local count
due to missing constraints is determined by evaluating the surface charge due to
the dipole moment of the surface unit cell:
$\nu_\text{L}=\mathbf{G}\cdot\Rl/2\pi$, where $\mathbf{G}$ is the reciprocal
lattice vector that points outward from the edge.

The second contribution is the topological count, which is determined by the
properties of the gapped bulk away from the edge. The form of the compatibility
matrix depends on the choice of unit cell; to calculate bulk properties such as
the winding numbers $n_i$, a natural choice is a balanced unit cell with
$\mathbf{d}=0$, such as the unit cells highlighted in the first rows of main
text Figs.~2 and 3. For lattices with one or more nonzero winding numbers $n_i$,
additional zero modes arise from the existence of the bulk topological invariant
$\Rt \neq 0$, which induces $\nu_\text{T}=\mathbf{G}\cdot\Rt/2\pi$ modes per
edge unit cell~\cite{Kane2014}. $\Rt$ is an extra dipole moment contribution
that is not apparent purely from geometric considerations of placing nodes and
links as in Eq.~\ref{eq:dipolemoment}, and is therefore termed a topological
polarization. For a section with two parallel edges such as in main text
Fig.~2d, $\mathbf{G}$ flips sign from one edge to the other whereas $\Rt$ is a
bulk property. The topological mode count is equal and opposite on the edges,
analogous to a transfer of ``charge'' (i.e. degrees of freedom) from one edge to
the other along the polarization direction.

We now calculate the zero mode count at each edge of the strip of the polarized
martini lattice in main text Fig.~2d) with periodic boundary conditions along
the vertical direction and free edges on the left and right. The left free edge,
which runs perpendicular to 
the reciprocal lattice vector $\mathbf{G}=-4\pi\hat{x}/a$, carries an edge
dipole moment $\Rl^1=-\mathbf{a}_1$. The total number of zero modes per unit
cell along that edge is $\mathbf{G}^1\cdot(\Rt+\Rl^{1})/2\pi=4$.
Fig.~2d (main text) shows one such local zero mode, numerically
calculated assuming periodic boundary conditions along the vertical direction.
Conversely, the right free edge has $\Rl^2=\mathbf{a}_1=-\Rt$, and the net mode
count is \emph{zero}. Although the nodes along the edge are missing links
relative to the interior, the local and topological contributions to the
unconstrained degrees of freedom cancel. This asymmetry in the linear zero mode
count translates to a drastic difference in rigidity between the left and right
edges of the prototype probed in Fig.~2e and Supplementary Movie 1. 

\section{Localization and size dependence of edge modes in finite samples} \label{app:edgenumerics}

\begin{figure*}
  \centering
  \includegraphics{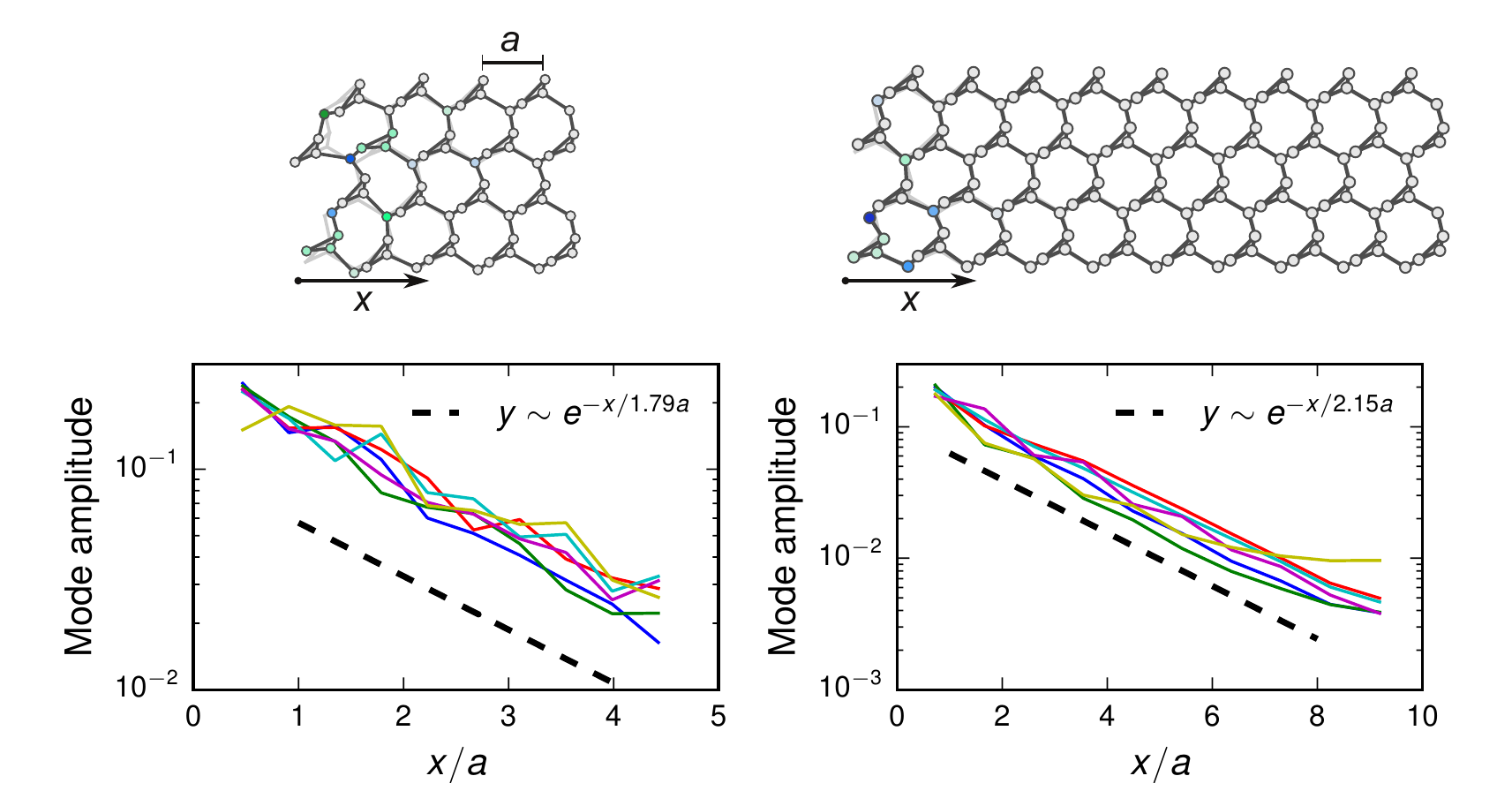}
  \caption{Decay of displacement amplitudes for non-translational zero modes as a
    function of distance from the left edge for systems with free boundaries
    perpendicular to the $x$ direction and periodic boundary conditions along $y$.
    Numerically, eight zero modes are found as predicted by the index theorem.
    Two of these are the uniform translations; the remaining six are shown to
    decay exponentially from the left edge with a decay length comparable to the
    lattice spacing $a$. The dashed line is a fit of the sum of the mode
    amplitudes to a simple exponential decay. Left, system shown in main text Fig.~2d.
    Right, system with twice the separation between the free left and right edges.
  } \label{fig:decay}
\end{figure*}

Appendix~\ref{app:pol} describes the local and topological count of zero modes
or states of self-stress exponentially localized to an edge of a semi-infinite
system. For example, the left edge of the strip shown in main text Fig.~2d is
predicted to have four zero modes per repeating unit in isolation, whereas the
right edge has no zero modes in isolation. In this situation, however, the
conclusions hold even for a strip of finite width, where the left and right
edges are separated by only a few lattice spacings. To understand this, we 
note that a zero mode is a set of displacements and rotations which does not
violate any link constraints. If this is true for a semi-infinite system, it is also true
for a subsection cut out of that system, because the displacements do not rely
on any stresses being borne by the cut links. Therefore, zero modes of
a semi-infinite system are also zero modes of the finite-width strip. Since all
the zero modes are localized to the left edge, they remain so even when the
right edge is brought close to the left edge. 

To illustrate this, we numerically study the spatial structure of the
topological zero modes for the strip in main text Fig.~2d, which are the zero
eigenvectors of the dynamical matrix $\mathbf{D=C^T C}$ (we set all constraint
stiffnesses/spring constants to 1). The free left and right edges are obtained
by removing four links from a periodic system, so the compatibility matrix has
eight fewer rows than columns. Its null space, which is also the null space of
$\mathbf{D}$, contains eight independent vectors which are the zero modes of the
structure. Two of these are rigid-body translations along $x$ and $y$; we remove
the corresponding displacement vectors from the null space to obtain six
independent null vectors with no translation component. Fig.~\ref{fig:decay}
shows the displacement amplitude averaged along the vertical direction as a
function of distance from the left edge for these six modes. All modes decay
exponentially as a function of distance from the left edge, with a decay length
of about two lattice spacings, and the mode amplitudes are diminished by an
order of magnitude at the right edge even for a strip that is just four lattice
units wide. The right edge does not acquire new zero modes in the finite system,
because the index theorem (Equation 1) constrains the total number of zero modes
minus stress states for the missing links, and all eight excess zero modes are
accounted for. The right edge could acquire new zero modes only of states of
self-stress arose elsewhere in the system to keep $\nm-\nss=8$ constant.

We note that the argument does not hold for states of self-stress: a state
of self-stress localized to an edge of a semi-infinite system is not
automatically a state of self-stress of a subsystem cut out of it, because it
might rely on some of the cut links to maintain equilibrium.

The physical prototype, main text Fig.~2e, does include additional constraints
because the top and bottom points have been pinned. Therefore, some of the zero
modes associated with the left edge will acquire a small stiffness.
Nevertheless, the additional constraints cannot make the right edge soft, and
the asymmetry of stiffness between left and right edges persists in the presence
of pinning, as is observed upon manually probing the edges (Supplementary Movie
1).

\section{Zero modes and states of self-stress at domain walls} \label{app:domainwall}

\begin{figure*}
  \centering
  \includegraphics{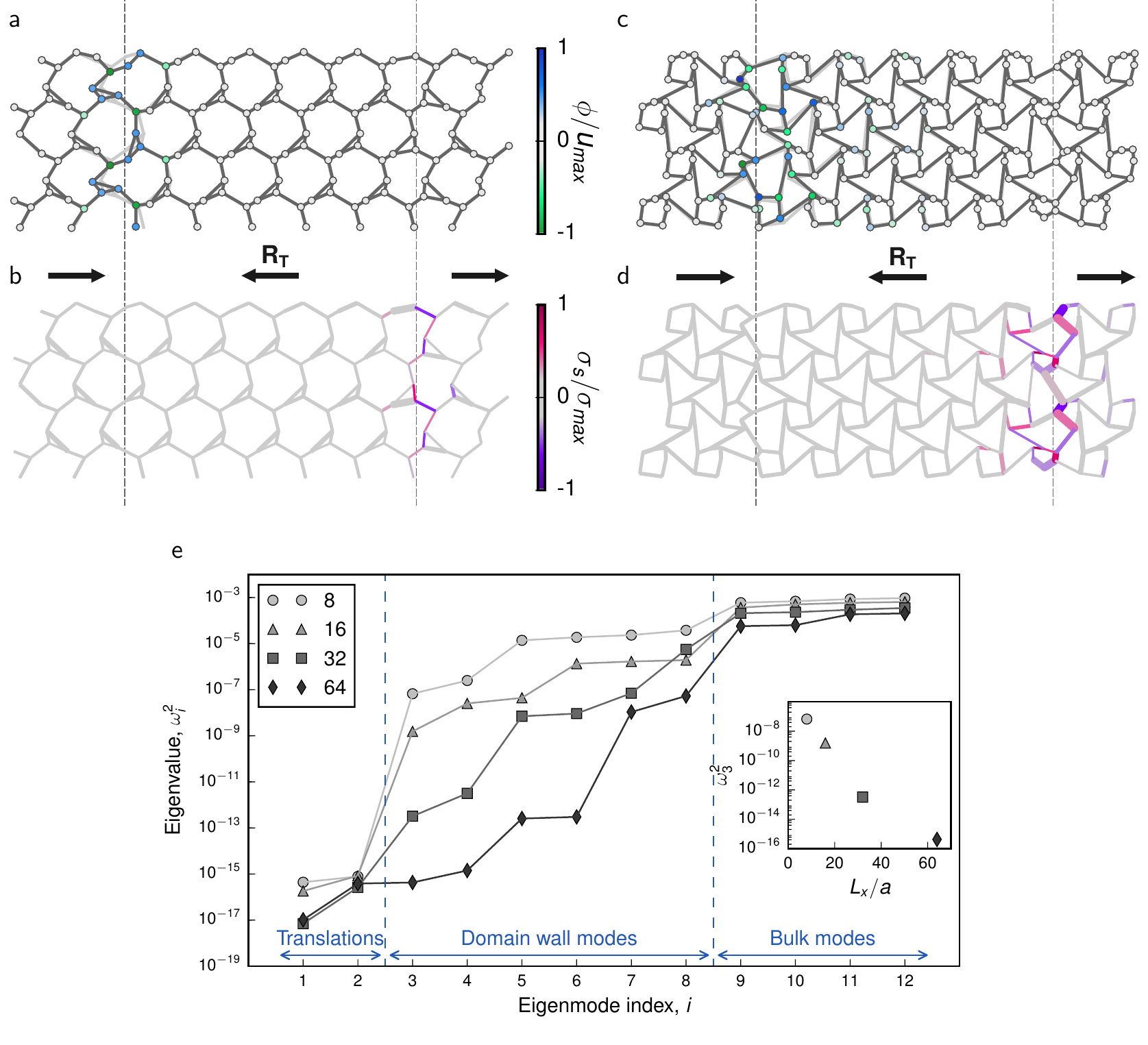}
  \caption{{\bfseries Topological modes and states of self-stress at domain
      walls.}
    {\bfseries a--b,} Domain walls (dotted lines) separating oppositely-oriented
    sections of the polarized martini lattice. Although the number of degrees and
    freedom and constraints are unchanged along the domain walls, they localize
    zero-energy modes and stress states because they carry a net polarization
    flux. The domain wall on the left has a net influx of polarization and localizes zero
    modes. In finite systems, these become soft modes and stress states with
    energy exponentially small in system size, one of which is visualized in
    {\bfseries a}. The dark lines show a
    lattice with nodes moved in proportion to the displacement components of the
    zero eigenvector, and coloured by the torsional components. The light grey lines
    show the undeformed lattice for comparison. The domain wall on the right has a
    net outflux of polarization and localizes states of self-stress, one of which is
    visualized in {\bfseries b}. Bonds are coloured by the shear components of
    the zero stress eigenvector, and their thicknesses are proportional to the
    tension components.
    {\bfseries c--d,} Same as in {\bfseries a--b}, for the polarized deformed
    hexachiral lattice.
    {\bfseries e,} System size dependence of the localized mode energies.
    The lowest eigenvalues of the dynamical matrix are shown as a
    function of system length along the horizontal direction for the domain wall
    geometry of panels {\bfseries a--b}, with domain walls always placed 0.66
    $L_x$ apart.  Inset: eigenvalue of the first non-translation mode (index
    $i=3$) as a function of system size.
  }
  \label{fig:domainwalls}
\end{figure*}

Besides influencing the count of zero modes at lattice edges, the topological
polarization also induces zero modes and states of self-stress at domain walls
separating lattices of different polarizations, even though no links have been
cut or added relative to a perfectly balanced (isostatic) system.
Ref.~\onlinecite{Kane2014} showed that, when a region $S$ of a system has no
missing bonds (i.e. all points satisfy the Maxwell criterion), then the
``topological count'' of the number of zero modes minus the number of states of
self-stress in the interior of the region is given by the net flux of
polarization through its boundary: in two dimensions, $\nu_\text{T}
=(1/A_\text{cell}) \int_{\partial S} \Rt \cdot \hat{n} dS$, where
$A_\text{cell}$ is the area of a unit cell, $\hat{n}$ is the inward boundary
normal of the boundary $\partial S$ which is required to lie in a region with a
spectral gap (i.e. there are no zero modes or states of self-stress on the
boundary of the region).

Consider the domain wall geometry in Fig.~\ref{fig:domainwalls}a, with periodic
boundary conditions along both directions so that every node has exactly three
links. Two domain walls (dashed lines) separate a region of the polarized martini lattice from
a region of the same lattice rotated by $\pi$ radians, which also rotates the
topological polarization. Therefore, a region surrounding the left domain wall
has a net influx of polarization, whereas a region surrounding the right domain
wall has a net outflux. Calculating the fluxes, we expect eight zero modes exponentially
localized to the left domain wall and eight states of self-stress localized to
the right domain wall. However, because the separation between the domain walls
is finite, the boundaries of each region do not lie in fully gapped parts of the
lattice because there are components of the topological modes from the other
domain wall in the same part, albeit exponentially small in the distance from
the other domain wall. Therefore, we expect corrections to the energies of the
localized modes that fall exponentially with system size. 

To test the dependence of the mode energies on system size, we numerically
computed the lowest eigenvalues of the dynamical matrix for systems with
different lengths ranging from $L_x = 8$ to 64 unit cells. In each case, the
domain wall separation was $0.66L_x$. The results (Fig.~\ref{fig:domainwalls}e)
show that the lowest eight eigenvalues have a qualitatively different system
size dependence than the higher eigenmodes. Rigid-body translations always make
up the lowest two modes (which are zero energy up to numerical precision) but
the next six modes have significant weights on the left domain wall. Above
eigenmode 9, the modes have the form of long-wavelength planewaves, delocalized
into the bulk. The energies of the domain wall modes fall sharply with system
size compared to the bulk modes, and at the larger system size the two
lowest-energy domain wall modes are indistinguishable in energy from the
rigid-body translations.  The inset shows the exponential fall in eigenvalue
with system size, consistent with the expected finite size correction to the
mode energy.

\section{Alternative link design} \label{app:alternative}

The gear network discussed in the main text deviates from spring network models.
Spring networks consist of freely pivoting nodes, connected by springs that
provide tensile constraints between them. By contrast, pivoting node motions are constrained in the geared network, so that only equal and opposite
rotations are allowed at each pair of adjacent nodes. We propose an alternative
link design that replicates this constraint on nodal rotations in the linear
regime, without the use of gears.

\begin{figure}
  \centering
  \includegraphics[]{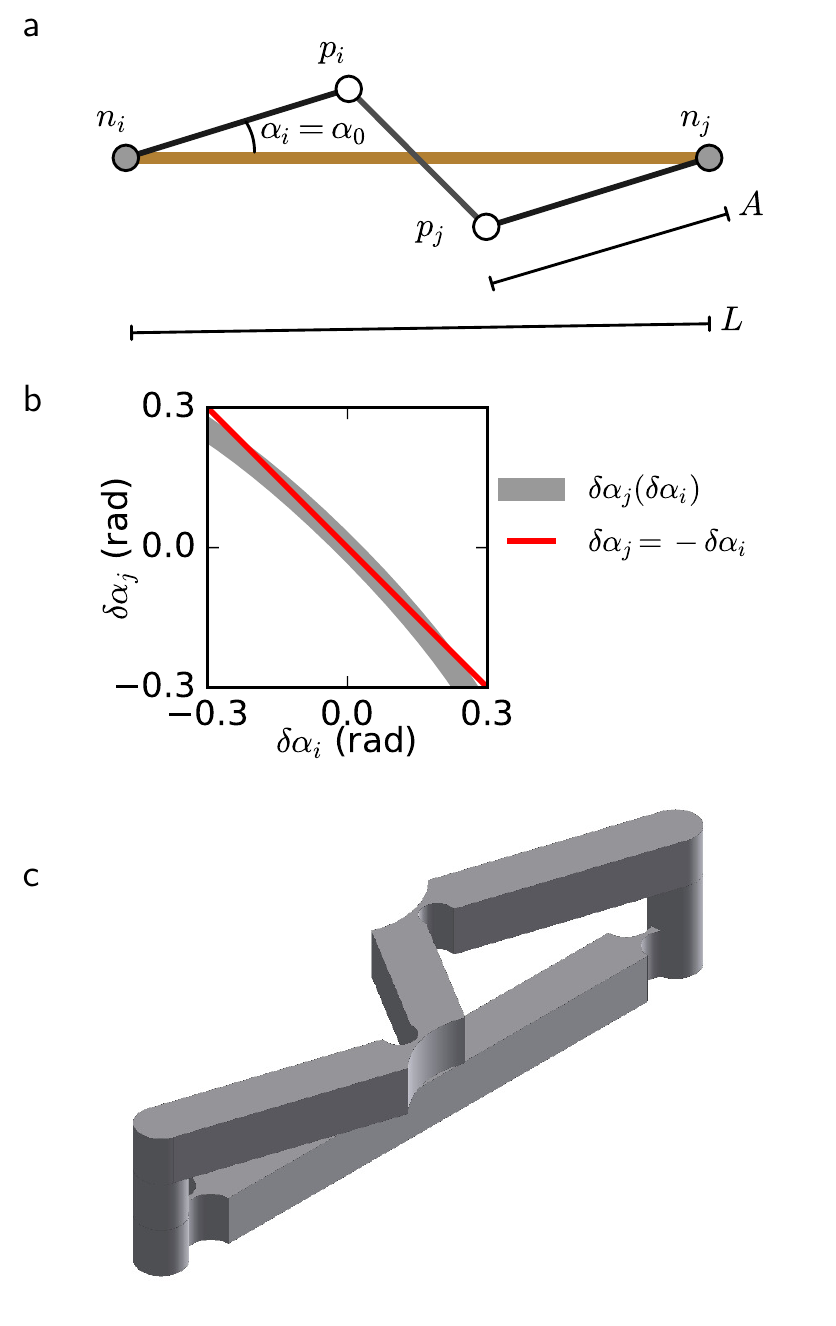}
  \caption{{\bfseries An alternative to the geared link.} {\bfseries a,} The
    link constrains the motions of freely pivoting nodes $n_i$,$n_j$ (grey
    discs); it is symmetric under in-plane rotations by $\pi$ radians. The nodes are
    connected by a rigid beam of length $L$ (orange line). Node $n_i$ is
    connected to subnode $p_i$ (white disc) by a rigid subbeam (black line) of
    length $A$ at an angle $\alpha_0$. Subnodes $p_i$,$p_j$ are interconnected by a
    rigid subbeam of length $l$ (grey line). The beam of length $L$ constrains the distance
    between nodes $n_i$,$n_j$, and the subbeam mechanism ensures that the nodes
    counterrotate in the linear regime. {\bfseries b,} Numerical calculation of
    the relationship between deviations $\delta \alpha_i,\delta \alpha_j$ from
    the angle $\alpha_0$ at nodes $n_i$,$n_j$ (grey line), for the link design
    shown in (a). Here, $\alpha_0 = 0.3 \mathrm{rad}$ and $A/L = 0.4$. In the
    linear regime, the relationship shows excellent agreement with $\delta
    \alpha_j = - \delta \alpha_i$ (red line), which describes pure
    counterrotation of the nodes. {\bfseries c,} A possible continuum 3D design
    for the link. Thin hinges accommodate low-energy pivoting at nodal points
    within the structure.}
  \label{fig:linkdesign}
\end{figure}

The alternative link design is shown in Fig.~\ref{fig:linkdesign}a. The bars in the link are
assumed to resist stretching, shearing and bending, while all nodes may pivot
freely.
In this linking mechanism, the polar angle $\alpha_i = \alpha_0$ at
node $n_i$ may be increased to an angle $\alpha_i = \alpha_0 + \delta \alpha_i$. As a result, the polar angle $\alpha_j = \pi + \alpha_0$ at node $n_j$ increases to $\alpha_j = \pi + \alpha_0 + \delta \alpha_j$. The relation
between $\delta \alpha_i$ and $\delta \alpha_j$ is given in the linear regime by
\begin{equation}\label{eq:proportional}
  \delta \alpha_j = -\delta \alpha_i,
\end{equation}
which is obtained as follows. Assuming no deformations in the beams, 
\begin{equation}
  |\vec{p_1} - \vec{p_2}|^2 = l^2,
\end{equation}
where $l$ is the length of the subbeam connecting subnodes $p_1$ and $p_2$,
holds for any angles $\alpha_i,\alpha_j$ compatible with the link's geometry.
Now, 
\begin{equation}
  \begin{split}
    \vec{p_1} &= A \cos(\alpha_i)\hat{x} + A \sin(\alpha_i)\hat{y}\\
    \vec{p_2} &= (L-A \cos(\alpha_j))\hat{x} -A \sin(\alpha_j)\hat{y};
  \end{split}
\end{equation}
setting 1) $\alpha_i = \alpha_j = \alpha_0$ and 2) $\alpha_i = \alpha_0 + \delta
\alpha_i$ and $\alpha_j = \alpha_0 + \delta \alpha_j$, we obtain a system of
equations that we can solve in a straightforward manner for $\delta \alpha_i$ and $\delta \alpha_j$ to
linear order. This calculation results in the relation of
Eq.~\ref{eq:proportional}, showing that the nodes in this mechanism purely
counterrotate in the linear regime. From numerical calculations, deviations
from this counterrotating behaviour are smaller than 5$\%$ up to $\delta
\alpha_i \sim 0.2\alpha_0$, at least where $0.25 < A/L < 0.65$ and $0.15
\mathrm{~rad } < \alpha_0 < 0.55 \mathrm{~rad }$ (Fig.~\ref{fig:linkdesign}b).
	
Note that for angle deviations $\delta \alpha$ outside of the linear regime, the linear
relation between nodal angles is no longer valid. In addition, if the nodal
angles are such that $\vec{n_i}$,$\vec{p_1}$ and $\vec{p_2}$ are collinear, the
mechanism may convert to one in which corotations (rather than counterrotations)
of the nodes are enforced.
In addition, the desired behaviour breaks down if the $L$-beam is
allowed to stretch significantly (similarly to the geared link). Deviations
$\delta L$ in the length $L$ are coupled to corotations of the nodes, or
shearing of the mechanism:
\begin{equation}
  \delta \alpha_i = \delta \alpha_j = \frac{2A-L}{2AL\alpha_0} \delta L.
\end{equation}

A possible continuum realization of the link design is shown in
Fig.~\ref{fig:linkdesign}c. This design may be used to build up a lattice with
many unit cells that forms a continuum solid, which can be manufactured via 3D
printing or other additive manufacturing techniques. In a network built up out
of such elements, all modes will cost a finite amount of energy due to bending
at thin hinges inside the link. States of self stress will exhibit themselves as
stretching, buckling or bending of these thin hinges rather than failure of the
thicker beams. Self-intersections can be avoided with layering and appropriate
choices of $A/L$ and $\alpha$.

\end{document}